\documentclass[nofootinbib,showpacs,preprint,aps]{revtex4-1}
\usepackage{graphicx}
\usepackage{dcolumn}
\usepackage{bm}

\usepackage{relsize}

\begin{document}


\title{Quark-lepton symmetric model at the LHC}

\author{Jackson D. Clarke}
\author{Robert Foot}%
\author{Raymond R. Volkas}
\affiliation{%
ARC Centre of Excellence for Particle Physics at the Terascale\\ School of Physics, University of Melbourne, 3010, Australia.}%

\date{\today}

\begin{abstract}
We investigate the quark-lepton symmetric model of Foot and Lew in the context of the Large Hadron Collider (LHC). 
In this `bottom-up' extension to the Standard Model, quark-lepton symmetry is achieved by introducing a gauged `leptonic colour' 
symmetry which is spontaneously broken above the electroweak scale. 
If this breaking occurs at the TeV scale, then we expect new physics to be discovered at the LHC. We examine 
three areas of interest: the Z$'$ heavy neutral gauge boson, charge $\pm1/2$ exotic leptons, and a colour triplet scalar diquark. 
We find that the LHC has already explored and/or will explore new parameter space for these particles over the course of its lifetime.
\end{abstract}

\maketitle

\section{Introduction}

The Large Hadron Collider (LHC) has begun probing physics at the TeV scale.  The goals are to discover the origin of
electroweak symmetry breaking and to search for new physics.  The purpose of this paper is to analyse the LHC
phenomenology of the discrete quark-lepton (q-l) symmetric model of Foot and Lew \cite{foot90}.  This theory
is conceptually related to both the Pati-Salam model \cite{patisalam} and the left-right symmetric model \cite{lrsymmetric}.  
Like the Pati-Salam model, it unifies the quantum numbers of the leptons with those of the quarks, but it does so using a
discrete exchange symmetry rather than a continuous symmetry.  This discrete symmetry can be viewed as the quark/lepton analogue
of the discrete (parity or charge-conjugation) symmetry used in the left-right symmetric model to unify the quantum numbers
of left-handed fermions with their right-handed partners.  The q-l symmetric model is distinct from grand unification, because it does
not require the concomitant unification of the strong and electroweak interactions.\footnote{It is, however, compatible with an eventual
high-scale grand unification, through a structure now known as ``quartification'' \cite{joshi91,*babu03,*chen04,*demaria05,*demaria05err,*demaria06,*demaria07,*babu08,*eby11}.}  
The absence of a continuous symmetry
transforming quarks into leptons and vice-versa allows the scale of the new physics to be as low as a TeV, making it testable
at the LHC.

We now review how the q-l symmetric model incorporates a discrete symmetry between quarks and leptons. Under the standard model
(SM) gauge group,
\begin{eqnarray}
	G_{SM}=SU(3)_q\times SU(2)_L\times U(1)_Y ,
\end{eqnarray}
quarks and leptons transform as
\begin{eqnarray}
\begin{array}{ll}
	Q_L\sim (3,2)(1/3),	&		f_L\sim (1,2)(-1),\\
	u_R\sim (3,1)(4/3),	&		e_R\sim (1,1)(-2),\\
        d_R\sim (3,1)(-2/3),   &              \nu_R \sim (1,1)(0) ,\\
\end{array}
\end{eqnarray}
where we have included a right-handed neutrino for neutrino mass generation purposes.
Some similarities between quarks and leptons are immediately obvious. 
They are both fermions in three generations and each sector transforms as $2 \oplus 1 \oplus 1$ under SU(2)$_L$. 
However, as is self-evident, quarks have a colour charge under SU(3)$_q$ while leptons do not, and the hypercharges are different.

One can increase quark-lepton symmetry by introducing a `leptonic colour' gauge interaction through an independent SU(3) group. 
The SM gauge group is enlarged to
\begin{eqnarray}
	G_{ql}=SU(3)_l\times SU(3)_q\times SU(2)_L\times U(1)_X , \label{Gql}
\end{eqnarray}
where SU(3)$_l$ is the new leptonic colour group and $X$ is a charge described in Section \ref{sec2}. 
The number of leptons is tripled, and SU(3)$_l$ must be spontaneously broken.
This is achieved  through one or more scalar multiplets that carry leptonic colour acquiring nonzero vacuum expectation values (VEVs).

After this breaking, the SM is recovered as a low-energy effective theory. The breaking pattern is:
\begin{eqnarray}
\begin{array}{c}
	SU(3)_l\times SU(3)_q\times SU(2)_L\times U(1)_X  \\
	\downarrow w \\
	SU(2)'\times SU(3)_q\times SU(2)_L\times U(1)_Y \\
	\downarrow u \\
	SU(2)'\times SU(3)_q\times U(1)_Q ,
\end{array} \label{breaking}
\end{eqnarray}
where $w$ and $u$ are the breaking scales, $Q$ is electric charge and SU(2)$'$ remains as an unbroken remnant of SU(3)$_l$. 
This SU(2)$'$ group is asymptotically free and expected to be confining.  Since all particles in non-trivial SU(2)$'$ representations 
are very massive (see below), the low-energy limit is indeed the SM despite this additional gauge force.

This paper is structured as follows: Section \ref{sec2} is an overview of the construction of the q-l symmetric model. We then move on to a study of phenomenology at the LHC. 
In Sec.~\ref{sec3} we look at a heavy neutral 
gauge boson called Z$'$. We use LHC data to put a lower bound on the mass of this particle. In Sec.~\ref{sec4} we 
investigate the charge $\pm1/2$ exotic leptons which gain a large mass at the leptonic colour breaking scale. We show that the LHC will 
explore new parameter space for these particles in the near future. In Sec.~\ref{sec5} we look at a colour triplet scalar diquark.
Using Tevatron and LHC data we constrain mass-coupling parameter space for this diquark. Section \ref{sec6} contains the conclusion.

\section{\label{sec2} Quark-lepton symmetric model}

We consider the q-l symmetric model constructed in Ref.~\cite{foot91}. When the SM gauge group is enlarged as in Eq.~(\ref{Gql}) we write the quark-lepton spectrum as
\begin{eqnarray}
\begin{array}{ll}
	Q_L\sim (1,3,2)(1/3),	 &		F_L\sim (3,1,2)(-1/3),\\
	u_R\sim (1,3,1)(4/3), &		E_R\sim (3,1,1)(-4/3),\\
  d_R\sim (1,3,1)(-2/3), &   N_R\sim (3,1,1)(2/3).\\
\end{array} \label{qlspectrum}
\end{eqnarray}
The leptonic components can be expressed as:
\begin{eqnarray}
\begin{array}{clr}
	F_L=\left(
		\begin{array}{clrrr}
			f_L		\\
			F_{1L}\\
			F_{2L}
		\end{array}
\right),
&
	E_R=\left(
		\begin{array}{clrrr}
			e_R		\\
			E_{1R}\\
			E_{2R}
		\end{array}
\right),
&
	N_R=\left(
		\begin{array}{clrrr}
			\nu_R		\\
			V_{1R}\\
			V_{2R}
		\end{array}
\right),
\end{array}
\end{eqnarray}
where $f_L$, $e_R$, $\nu_R$ are the familiar SM leptons and right-handed neutrino, and $F_{iL}$, $E_{iR}$, $V_{iR}$ ($i=1,2$) are the exotic fermions named `liptons'. 
The generational indices are suppressed. The components of $F_L$ are each doublets under SU(2)$_L$:
\begin{eqnarray}
\begin{array}{clr}
	f_L=\left(
		\begin{array}{clrr}
			\nu_L		\\
			e_L
		\end{array}
\right),
&
	F_{iL}=\left(
		\begin{array}{clrr}
			V_{iL}		\\
			E_{iL}
		\end{array}
\right).
\end{array}
\end{eqnarray}

The simplest way to achieve electroweak symmetry breaking is through a Higgs doublet, $\phi\sim(1,1,2)(1)$, analogously to the SM. 
The simplest Yukawa Lagrangian,
\begin{eqnarray}
  \mathcal{L}^{(1)}_{Yuk}=\lambda_1\left[\overline{F}_LE_R\phi+\overline{Q}_Lu_R\phi^c\right]+\lambda_2\left[\overline{F}_LN_R\phi^c+\overline{Q}_Ld_R\phi\right]+H.c. ,
  \label{Yuk1}
\end{eqnarray}
where $\phi^c\equiv i\tau_2\phi^\ast$ and $\tau_2$ is the second Pauli matrix, leads to unacceptable tree-level mass relations between quarks and leptons. 
These relations are easily avoided by instead introducing two Higgs doublets, $\phi_1$ and $\phi_2$, as described in Ref.~\cite{levin93}.

Leptonic colour symmetry breaking is achieved by introducing the Higgs triplets $\chi_1\sim (3,1,1)(2/3)$ and $\chi_2\sim (1,3,1)(-2/3)$ through the Yukawa Lagrangian
\begin{eqnarray} \mathcal{L}^{(2)}_{Yuk}=h_1\left[\overline{(F_L)^c}F_L\chi_1+\overline{(Q_L)^c}Q_L\chi_2\right] +h_2\left[\overline{(E_R)^c}N_R\chi_1+\overline{(u_R)^c}d_R\chi_2\right]+H.c.
  \label{Yuk2}
\end{eqnarray}
The leptonic colour symmetry is broken by a nonzero VEV for $\chi_1$. It induces the breaking
\begin{eqnarray}
  SU(3)_l\times U(1)_X\longrightarrow SU(2)'\times U(1)_Y ,
\end{eqnarray}
where 
\begin{eqnarray}
  Y=X+\frac{1}{3}T
\end{eqnarray}
and $T=\text{diag}(-2,1,1)$. The $\chi_1$ VEV also generates masses for the liptons.

One can generate small $\nu$ masses in a number of ways, for example by inducing a see-saw mechanism through Higgs multiplets 
$\Delta_1\sim (\bar{6},1,1)(-4/3)$ and $\Delta_2\sim (1,\bar{6},1)(4/3)$ as in Refs.~\cite{foot91,levin93},  through additional Higgs doublets \cite{levin93},
or via additional singlet fermions \cite{newimp}.

One can now see that a $Z_2$ discrete symmetry of the Lagrangian 
\begin{eqnarray}
\begin{array}{lll}
	Q_L \leftrightarrow F_L,	  & \chi_1\leftrightarrow\chi_2 , &		G_q^\mu\leftrightarrow G_l^\mu , \\
	u_R \leftrightarrow E_R,	    & \phi\leftrightarrow\phi^c , &		C^\mu\leftrightarrow -C^\mu . \\
  d_R \leftrightarrow N_R, &      & 
\end{array}
\end{eqnarray}
can be postulated, where $G_{q,l}^\mu$ are the gauge bosons of SU(3)$_{q,l}$ and $C^\mu$ is the gauge boson of U(1)$_X$. This is the quark-lepton symmetry.

\section{\label{sec3}The Z$'$ neutral gauge boson}

\subsection{Coupling}

Like many extensions to the SM, the q-l symmetric model predicts a heavy extra neutral gauge boson, the Z$'$. The relevant neutral current part of the covariant derivative is
\begin{eqnarray}
  D^\mu_{n.c.} = \partial^\mu+i\frac{g_2}{2}W_3^\mu\tau_3+i\frac{g'}{2}C^\mu X+i\frac{g_s}{2\sqrt{3}}G_8^\mu T , \label{partiald}
\end{eqnarray}
where once again $T=\text{diag}(-2,1,1)$. Defining the VEVs~\footnote{For simplicity, we will write subsequent
equations for the case where only one electroweak Higgs doublet is
introduced. The generalisation to the realistic two-doublet situation 
does not affect any of the phenomenology we consider in the
rest of this paper.} 
$\left\langle \phi\right\rangle \equiv u$ and $\left\langle\chi_1\right\rangle \equiv w$, the neutral gauge boson mass matrix is
\begin{eqnarray}
  M^2 = \left(
  \begin{array}{cccrrr}
    g_2^2u^2/2		&		-g_2g'u^2/2							&		0 \\
    -g_2g'u^2/2		&		g'^2u^2/2+2g'^2w^2/9		&		-2\sqrt{3}g'g_sw^2/9 \\
    0							&		-2\sqrt{3}g'g_sw^2/9		&		2g_s^2w^2/3 
  \end{array}
  \right) \label{m2matrix}.
\end{eqnarray}
Solving the eigenvalue problem gives
\begin{eqnarray}
	M^2_\gamma  &=&  0 , \nonumber \\
	M^2_{\stackrel{Z}{\mathsmaller{\mathsmaller{Z'}}}}  &=& u^2\kappa_1+w^2\kappa_2 \mp	\sqrt{\left(u^2\kappa_1+w^2\kappa_2\right)^2-u^2w^2\kappa_3},
\end{eqnarray}
where the minus (plus) sign corresponds to the Z (Z$'$) and
\begin{eqnarray}
\begin{array}{lcl}
	\kappa_1 &=& \frac{1}{4}\left(g_2^2+g'^2\right),  \\
	\kappa_2 &=& \frac{1}{9}\left(g'^2+3g_s^2\right), \\
	\kappa_3 &=& \frac{1}{9}\left(g_2^2g'^2+3g_2^2g_s^2+3g'^2g_s^2\right) . 
\end{array}
\end{eqnarray}
The SM couplings are related to the q-l symmetric couplings by
\begin{eqnarray}
  \sin^2{(\theta_W)_{SM}} &\equiv & \frac{e^2}{g_2^2} = \frac{g'^2g_s^2}{3\kappa_3}, \nonumber \\
  {g'_{SM}}^2 &\equiv & {e^2 \over \cos^2 (\theta_W)_{SM}} = \frac{g'^2g_s^2}{3\kappa_2}.
\end{eqnarray}
Furthermore we have
\begin{eqnarray}
  u^2&=&\frac{2M_W^2}{g_2^2}, \nonumber \\
  w^2&=&\frac{g_2^2}{2\kappa_3}\frac{M_Z^2M_{Z'}^2}{M_W^2} . \label{wdef}
\end{eqnarray}

%

The neutral current covariant derivative (\ref{partiald}) is rewritten in the gauge-boson mass eigenstate basis as
\begin{eqnarray}
  D^\mu_{n.c.} = \partial^\mu+iQ_\gamma A^\mu+iQ_ZZ^\mu+iQ_{Z'}Z'^\mu ,
\end{eqnarray}
where $Q_\gamma$, $Q_Z$ and $Q_{Z'}$ are the charges that couple to $A^\mu$, $Z^\mu$ and $Z'^\mu$, respectively. 
The charges coupling to Z and Z$'$ are given by
\begin{eqnarray}
  Q_{\stackrel{Z}{\mathsmaller{\mathsmaller{Z'}}}}
  &=& \pm\left(\frac{\sin^2{(\theta_W)_{SM}}}{{g'_{SM}}^2}+\frac{1}{4\kappa_1}\left[\eta_{\stackrel{Z}{\mathsmaller{\mathsmaller{Z'}}}}^2-\frac{g'^4}{9\kappa_3}\right]\right)^{-1/2} \nonumber \\
  & & \times \left(
      \begin{array}{l}
        \frac{\tau_3}{2}-\left(\sin^2{(\theta_W)_{SM}}-\frac{g_2^2g'^2}{12\kappa_1\sqrt{\kappa_3}}\left[\eta_{\stackrel{Z}{\mathsmaller{\mathsmaller{Z'}}}}-\frac{g'^2}{3\sqrt{\kappa_3}}\right]\right)Q \\
        - \left(\frac{\sqrt{\kappa_3}}{8\kappa_1}\left[\eta_{\stackrel{Z}{\mathsmaller{\mathsmaller{Z'}}}}-\frac{g'^2}{3\sqrt{\kappa_3}}\right]\right)T
      \end{array}
      \right) ,
  \label{qzzpcoupling}
\end{eqnarray}
where
\begin{eqnarray}
  \eta_{\stackrel{Z}{\mathsmaller{\mathsmaller{Z'}}}} 
    = \left(\frac{3\sqrt{\kappa_3}}{g'^2}-\frac{6\kappa_1}{g'^2\sqrt{\kappa_3}}\frac{M^2_{\stackrel{Z}{\mathsmaller{\mathsmaller{Z'}}}}}{w^2}\right)^{-1} .
\end{eqnarray}
In the $w\rightarrow\infty$ limit the quantities within square brackets in Eq. (\ref{qzzpcoupling}) tend to zero for the charge coupling to Z, so that the expression reduces to the SM coupling. To facilitate the calculation of these couplings we write the quantum numbers of the fermions in Table \ref{qnumbers}.


\begin{table*}[t]
	\begin{ruledtabular}
	\begin{tabular}{ccccccccccccc}
    &$\nu_L$&$\nu_R$&$e_L$&$e_R$&$V_{iL}$&$V_{iR}$&$E_{iL}$&$E_{iR}$&$u_L$&$u_R$&$d_L$&$d_R$ \\ \hline
    $T$&$-2$&$-2$&$-2$&$-2$&1&1&1&1&0&0&0&0\\
    $Q$&0&0&$-1$&$-1$&$+1/2$&$+1/2$&$-1/2$&$-1/2$&$+2/3$&$+2/3$&$-1/3$&$-1/3$ \\
    $\tau_3$&$+1$&0&$-1$&0&$+1$&0&$-1$&0&$+1$&0&$-1$&0 \\ 
	\end{tabular}
  \end{ruledtabular}
\caption{\label{qnumbers} The quantum numbers of the fermions.}
\end{table*}

\subsection{Mass limit}

A typical signature of a Z$'$ gauge boson is a high mass resonance in dilepton production through the Drell-Yan process 
$pp\stackrel{Z'}{\longrightarrow}l^+l^-$ \cite{drellyan71}, in both the dielectron and dimuon channels \cite{cdfdimuon,*d0dielectron,atlaszpnew,*cmsdilepton}. 
Recent approaches combine data from each channel to obtain a stronger bound than is possible with a single channel \citep{atlaszpnew,*cmsdilepton}.

The aim now is to use LHC data to put a constraint on the mass of the q-l symmetric Z$'$, henceforth called Z$'_{ql}$. 
We achieve this by calculating the cross section $\sigma(pp\stackrel{Z'}{\longrightarrow}l^+l^-)$, where $l$ can be either $\mu$ or $e$. 
As a comparison we also carry out this calculation for the benchmark model of a Z$'$ with SM coupling, henceforth called Z$'_{SM}$.

We start with the Z$'$ interaction Lagrangian,
\begin{eqnarray}
  \mathcal{L}_{Z'}=\bar{f}\gamma_\mu\left({g'}^f_V-{g'}^f_A\gamma_5\right)fZ'^\mu \label{nclag} ,
\end{eqnarray}
where
\begin{eqnarray}
  \begin{array}{lcl}
    {g'}^f_V&=&\frac{1}{2}Q_{Z'}(f_L)+\frac{1}{2}Q_{Z'}(f_R) , \\
    {g'}^f_A&=&\frac{1}{2}Q_{Z'}(f_L)-\frac{1}{2}Q_{Z'}(f_R) ,
  \end{array}\label{gzvaqz}
\end{eqnarray}
and $Q_{Z'}$ is the charge coupling to the Z$'$ boson. 

We use the Drell-Yan cross section,
\begin{eqnarray}
	\sigma(pp\rightarrow l^+l^-)&= &
	\sum_{q}\int_0^1\int_0^1dx_1dx_2
	\left[
	\begin{array}{l}
					f_{q/p}(x_1,Q^2)f_{\bar{q}/p}(x_2,Q^2) \\
					+ f_{q/p}(x_2,Q^2)f_{\bar{q}/p}(x_1,Q^2)
  \end{array}
  \right] \nonumber \\
  && \times
  \sigma(q\bar{q}\rightarrow l^+l^-;s=x_1x_2S_{beam}) , \label{drellyan}
\end{eqnarray}
for a $pp$ collision at $\sqrt{S_{beam}}=7$ TeV. The parton distribution functions (PDFs), $f_{q/p}(x,Q^2)$, are evaluated using the MSTW08 PDF set at $Q^2=\frac{1}{4}s$ \citep{mstw08}. 
We then scale the results for next-to-leading order effects using a $k$-factor of 1.3 \citep{hamberg91,*hamberg91b,*cmswp}.

This cross section is dominated by the parton processes $u\bar{u}\rightarrow l^+l^-$ and $d\bar{d}\rightarrow l^+l^-$. The parton cross sections in the $\sqrt{s}\gg m_q, m_{l}$ limit are given by
\begin{eqnarray}
  \sigma(q\bar{q}\stackrel{Z'}{\rightarrow} l^+l^-)=
     \frac{1}{3}\frac{1}{12\pi}\frac{s}{(s-M_{Z'}^2)^2+M_{Z'}^2\Gamma_{Z'}^2} \left(|{g'}^q_V|^2+|{g'}^q_A|^2\right)\left(|{g'}^{l}_V|^2+|{g'}^{l}_A|^2\right), \label{sigffmumu}
\end{eqnarray}
where $M_{Z'}$ is the mass of the Z$'$ boson. The Z$'$ width is given to a good approximation by
\begin{eqnarray}
	\Gamma_{Z'}&=&3\left(\Gamma(Z'\rightarrow \nu_e\bar{\nu_e})
						+ \Gamma(Z'\rightarrow e^+e^-) + \Gamma(Z'\rightarrow u\bar{u})
						+ \Gamma(Z'\rightarrow d\bar{d})\right) , \nonumber \\
  \Gamma(Z'\rightarrow f\bar{f})&=&N_c\frac{M_{Z'}}{12\pi}\left(|{g'}_V^{f}|^2+|{g'}_A^{f}|^2\right), \label{zfwid} 
\end{eqnarray}
where $N_c$ is a colour factor equal to 1 for leptons and 3 for quarks. 
For Z$'_{ql}$ we use Eqs.~(\ref{qzzpcoupling}) and (\ref{gzvaqz}) to find the couplings, 
whilst for Z$'_{SM}$ we use the familiar SM couplings.

\begin{figure}[t]
	\includegraphics{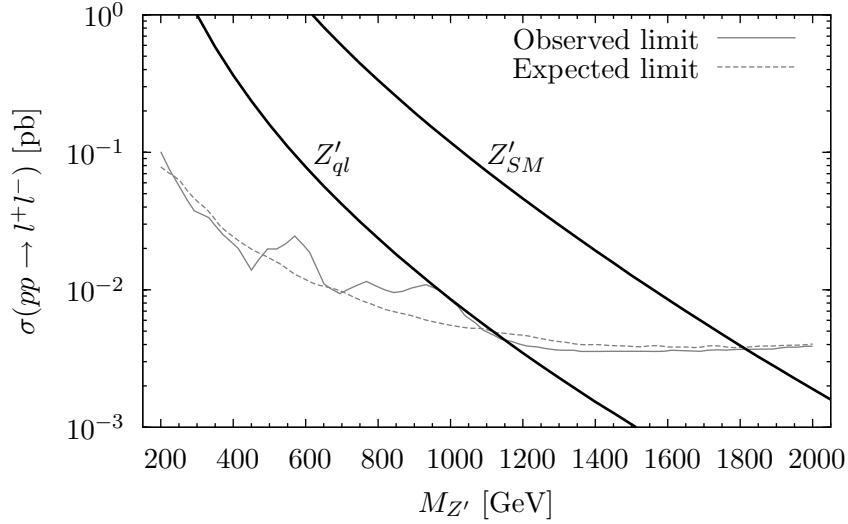}
	\caption{The cross section $\sigma(pp\rightarrow l^+l^-)$ versus $M_{Z'}$ for Z$'_{ql}$ and Z$'_{SM}$. 
Also shown are the $95\%$ confidence level upper limits on the cross section from ATLAS data \cite{atlaszpnew}.}
	\label{sigmamzp3}
\end{figure}

\begin{table}[t]
  \begin{ruledtabular}
	\begin{tabular}{lcc}
              & \multicolumn{2}{c}{Observed $M_{Z'}$ limit (GeV)} \\ 
              & Tevatron & LHC \\ \hline
		$Z'_{SM}$ & 1060               & 1820 \\
		$Z'_{ql}$ & 780                & 950 \\ 
	\end{tabular}
	\end{ruledtabular}
	\caption{\label{zpbounds} The observed 95$\%$ confidence level lower mass limits on Z$'_{SM}$ and Z$'_{ql}$.}
\end{table}

The results from our calculations are shown in Fig.~\ref{sigmamzp3}. The lower bounds on the $M_{Z'}$ parameters, 
listed in Table \ref{zpbounds}, 
are given by the upper limit on the cross section at the LHC \cite{atlaszpnew,*cmsdilepton}. We use the ATLAS limit, but a similar result is found using the CMS limit. 
The lower bound of 1820 GeV for Z$'_{SM}$ agrees with that in the literature. 
The lower bound of 950 GeV for Z$'_{ql}$ is an improvement of 230 GeV on the previous bound \cite{foot91}, and the best bound to date on a q-l symmetric Z$'$. 
The bound for Z$'_{ql}$ is smaller than that for Z$'_{SM}$ due to the quarkphobic nature of Z$'_{ql}$. In fact, $Br(Z'_{ql}\rightarrow \sum q\bar{q})$ is less than $0.5\%$.

An analogous calculation was performed for the Tevatron using the upper limit on the cross section from D0 \cite{d0dielectron}. 
The results are also given in Table \ref{zpbounds}. The LHC bounds already exceed those of the Tevatron. In the coming years, 
with more data and higher energy collisions, this bound is set to increase significantly if a Z$'$ is not discovered first.

In our calculations we have assumed that all the liptons are heavier than $\frac{1}{2}M_{Z'}$, so that none of them is a Z$'$ decay product. As will be shown in the following section, this need not be the case. We merely make the assumption for simplicity, in order not to introduce too many unknown variables. If liptons are present at low enough masses the Z$'$ width increases and the lower bound falls. As an example, for the case of a single left sector lipton of mass 200 GeV, the lower bound on the Z$'_{ql}$ mass from the LHC becomes 930 GeV.

\section{\label{sec4}Liptons}

\subsection{Mass eigenstates}

The left and right components of the liptons do not combine to form a Dirac fermion in the obvious way (i.e.\ $f\neq E_{iL}+E_{iR}$). 
To understand what actually happens, we need to examine the Lagrangian of Eq.~(\ref{Yuk2}); it has the form
\begin{eqnarray} 
  \mathcal{L}_{mass}=\overline{X}_L\mathcal{M}X_R+H.c. ,
\end{eqnarray}
where
\begin{eqnarray} 
  \begin{array}{cc}
  X_L=\left(
    \begin{array}{c}
      V_L \\
      (E_R)^c\\
    \end{array}
  \right) ,
  &
  X_R=\left(
    \begin{array}{c}
      (E_L)^c \\
      V_R\\
    \end{array}
  \right),
\end{array}
\end{eqnarray}
and
\begin{eqnarray} 
  \mathcal{M}=\left(
    \begin{array}{cc}
      m_1 & m_\nu \\
      m_e & m_2   \\
    \end{array}
    \right).
\end{eqnarray}
This is for the one-generation case with SU(2)$'$ indices suppressed. After symmetry breaking, written explicitly, 
the entries are $m_\nu=\lambda_2u$, $m_e=\lambda_1u$ (the usual neutrino and electron mass terms), 
$m_1=2h_1w$ and $m_2=h_2^\ast w$. 
The non-observation of liptons implies that the latter two terms are much larger than the former. We assume they are large 
enough so that the off-diagonal terms can be approximated as zero. In this case we get four exotic Dirac fermions per generation:
\begin{eqnarray} 
  f_{l1}& = &(E_{2L})^c+V_{1L},\nonumber \\
  f_{l2}& = &(E_{1L})^c+V_{2L}, \nonumber \\
  f_{r1}& = &(E_{2R})^c+V_{1R}, \nonumber \\
  f_{r2}& = &(E_{1R})^c+V_{2R}.  
   \label{diracliptons}
\end{eqnarray}
We call these left ($f_{lj}$) and right ($f_{rj}$) sector liptons, with masses $m_1$ and $m_2$ respectively. 
The liptons are degenerate within a sector, each having an SU(2)$'$ `colour' $j={1,2}$.  
The conjugation flips the handedness of the spinor, so we always have a left handed and right handed part in the Dirac fermion.

Of course we have little idea of the mass spectrum of these liptons. Our model does not predict the masses, 
just as the SM does not predict the quark and lepton masses. We only know that they are massive enough to have not yet been detected.

\subsection{Liptonium}

Liptons can be pair produced at the LHC through the Drell-Yan process. The cross section depends primarily on the mass of the liptons and 
moderately on the mass of the Z$'$. In the following analysis we focus on the lightest lipton, $L$. 
This lipton can be either from the left or the right sector, and we consider both possibilities.

Liptons are charged under the confining SU(2)$'$ group, so one might think that pair produced liptons will act like pair produced heavy quarks. In fact, because there are no light liptons to facilitate SU(2)$'$ `hadronisation', they act very differently. The pair, instead, forms an 
SU(2)$'$-neutral bound state, $L\bar{L}$, which we call `liptonium'. As we will see, energy is first 
lost through SU(2)$'$ glueball radiation. To avoid confusion, and in order to distinguish them from SU(3)$_q$ gluons, 
we will call the SU(2)$'$ exotic gluons `huons'. Of course, liptons are not charged under SU(3)$_q$ so do not couple to gluons.

Much of our expectations can be inferred from the work of Carlson et al.\ \cite{carlson91}, wherein the phenomenology and collider signatures of the colour SU(5) model 
of Foot and Hernandez \cite{foothernandez} were studied. 
This model has the same intermediate and low-energy gauge structure (\ref{breaking}) as the q-l symmetric model. The analogy to 
liptonium in the colour SU(5) model is quirkonium, formed from `quirks' which have the same quantum numbers as our liptons. 
These quirks are doublets under an unbroken group analogous to our SU(2)$'$ group.

Following Ref.~\cite{carlson91}, we assume that the invariant mass of the lipton pair is well above the threshold $\sim2m_L$, but below $4m_L$. 
The lipton pair can be viewed as a highly excited liptonium state. The de-excitation of this state will occur in two distinct stages.

The first stage is hueball emission. As the liptons oscillate in the potential, many hueballs will be radiated with no preferential 
direction in the liptonium frame. This will lead to a large amount of missing energy but only a small amount of missing transverse energy. 
The hueballs will also carry away a significant amount of angular momentum, leaving the liptonium with large $J$. 
When the excitation energy drops beneath the hueball mass, de-excitation will occur through photon emission. 
It is unclear whether these photons will be hard or soft compared to the hueball mass.

The second stage occurs when the liptons enter the one-huon-exchange Coulomb potential 
\begin{equation}
V(r)=-3\alpha_s(r)/4r, \label{onehuonpot}
\end{equation}
regime.
Here we assume that the liptonium has a large orbital quantum number $l$ and that spin states $S=1$ and $S=0$ are populated in the ratio $3:1$. 
Annihilation of the liptons to huons is small for high $l$ states, and $|\Delta l|=1$ electromagnetic transitions are favoured. 
(Spin flip transitions are small and can be ignored.) There will be many of these soft photon emissions until the liptonium is in the $n^3P$ or $n^1P$ state. 
From here, parity symmetry prevents all lipton pairs annihilating to invisible hueballs. The $n^1P$ states cascade predominantly to $n^1S$ states 
which annihilate to photons or huons. The $n^3P$ states can annihilate to huons or cascade to $n^3S$ states. 
The cascade always dominates \cite{carlson91}. The parity symmetry then prevents the $n^3S$ states annihilating entirely to huons, 
and $s$-channel neutral current annihilations are expected to dominate.

Thus we expect that approximately 75$\%$ of the total pair-produced liptons will eventually reach $^3S$ states. From here they can annihilate 
to $W^+W^-$, Z Higgs, $hh\gamma$, $hh$Z and $f\bar{f}$, where $h$ are huons. The most 
exciting products are SM fermions. Production through this mechanism will result in lepton pairs or jets of invariant mass $\sim2m_L$. 
Similarly, we expect that approximately 25$\%$ of pair-produced liptons will reach $^1S$ states where they can annihilate to invisible huons or diphotons.

If the decay of liptonium occurs as we expect then liptons with mass less than $\sim100$ GeV are ruled 
out by the Large Electron-Positron (LEP) collider experiments. This is because LEP, running at up to $\sqrt{S_{beam}}=209$ GeV, 
was able to tag large missing energy events \cite{lepmissinge}. If $m_L<100$ GeV then the large missing energy 
resulting from the first stage of liptonium de-excitation would have been detected at LEP in dilepton events. 
The LHC experiment does not have this capability, instead only being able to tag large missing transverse energy events.

\subsection{Seeing liptons at the LHC}

Our aim now is to determine the cross section $\sigma(pp\rightarrow f\bar{f})$ through the liptonium resonance. This cross section can be approximated by
\begin{eqnarray}
  \sigma(pp\stackrel{L\bar{L}}{\longrightarrow} l^+l^-)
     &\approx&\sigma(pp\rightarrow L\bar{L}) \times0.75\times Br(^3S_1\rightarrow l^+l^-) , \label{ppLLbarll}
\end{eqnarray}
where the factor of 0.75 accounts for the fraction of liptonium states which reach the $^3S_1$ state. The $\sigma(pp\rightarrow L\bar{L})$ 
and $Br(^3S_1\rightarrow l^+l^-)$ factors will be given in the following two subsections.

\subsubsection{$\sigma(pp\rightarrow L\bar{L})$}

In our analysis we ignore the contribution of the Z$'$ neutral current by taking the limit $w\rightarrow\infty$, where $Q_Z=Q_Z^{SM}$. 
This will be justified in a moment. We start with the neutral current interaction Lagrangian,
\begin{eqnarray}
  \mathcal{L}_{n.c.}= e\bar{f}\gamma_\mu QfA^\mu 
  + \bar{f}\gamma_\mu\left(g^f_V-g^f_A\gamma_5\right)fZ^\mu \label{nclag2} .
\end{eqnarray}
We calculate $\sigma(pp\rightarrow L\bar{L})$ using the Drell-Yan cross section (\ref{drellyan}). The parton cross sections are found in the $\sqrt{s},m_{L}\gg m_q$ limit using
\begin{eqnarray}
  \sigma(q\bar{q}\stackrel{\gamma,Z}{\longrightarrow} L\bar{L}) & = & 
  \frac{2}{3}Q_q^2Q_L^2\frac{8\pi\alpha^2}{s^2}\sqrt{1-\frac{4m_{L}^2}{s}} \nonumber \\
    &&\times\left\{ 
    \begin{array}{c}
    \left(1+2\textbf{Re}(\chi)g^q_Vg^{L}_V\right)\left[
      \begin{array}{c}
         \frac{1}{6}s
        +\frac{1}{3}m_{L}^2
      \end{array}\right]\\
    +\left|\chi\right|^2\left\{ 
    \begin{array}{c} \left(|g^q_V|^2+|g^q_A|^2\right)\left(|g^{L}_V|^2+|g^{L}_A|^2\right)\left[
      \begin{array}{c}
         \frac{1}{6}s
        -\frac{1}{6}m_{L}^2
      \end{array}\right]  \\
      +\frac{1}{2}\left(|g^q_V|^2+|g^q_A|^2\right)\left(|g^{L}_V|^2-|g^{L}_A|^2\right)m_{L}^2 \\
    \end{array}
    \right\} \\
  \end{array}
  \right\},  \label{sigmafdfdff}
\end{eqnarray}
where
\begin{equation}
	\chi=\frac{1}{4\pi Q_qQ_{L}\alpha}\frac{s}{(s-M_Z^2+iM_Z\Gamma_Z)}. \label{chidef}
\end{equation}
The $\chi$-independent term is the pure photon contribution, the $|\chi|^2$ term is the pure Z contribution, and the Re$(\chi)$ term is the $\gamma/Z$ interference term. 
The width, $\Gamma_Z$, is given by equations analogous to (\ref{zfwid}).

To justify ignoring the Z$'$ contribution, we consider the ratio
\begin{equation} 
  R = \sigma(q\bar{q}\stackrel{Z'}{\longrightarrow}L\bar{L})/\sigma(q\bar{q}\stackrel{\gamma,Z}{\longrightarrow}L\bar{L})
\end{equation}  
as a function of Z$'$ mass. The quantity $\sigma(q\bar{q}\stackrel{Z'}{\longrightarrow}L\bar{L})$ is found analogously to the pure Z 
contribution in Eq.~(\ref{sigmafdfdff}), with the width of Z$'$ modified to include decay to a single family of lightest 
liptons when $M_{Z'}>2m_L$. 
We find that $R$ is largest for left (right) sector liptons when $M_{Z'}\approx2.5m_L$, where $R\approx0.12$ $(R\approx0.56)$. Away from this resonance the contribution is much smaller, 
and given that $M_{Z'}>950$ GeV this coincidence is unlikely in the region that will be of interest.

\subsubsection{$Br(^3S_1\rightarrow l^+l^-)$}

\begin{figure}[t]
\includegraphics{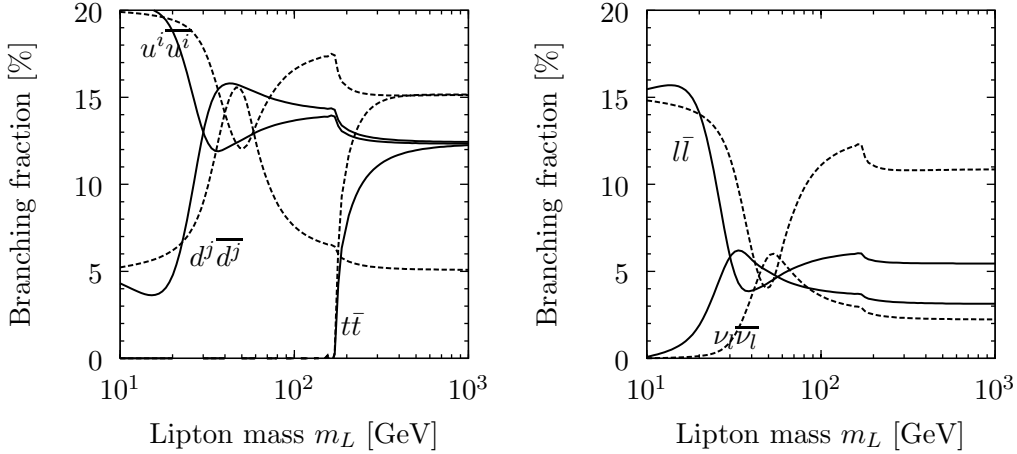}
\caption{\label{branchings} Branching fractions of the $^3S_1$ liptonium state to difermions. The solid (dashed) lines correspond to left (right) sector 
liptons and $i=\{1,2\}$, $j=\{1,2,3\}$ are family indices.}
\end{figure}

The decay rate of the $^3S_1$ liptonium state to a pair of fermions is given by the spin- and colour-summed squared matrix element multiplied by the square of the 
liptonium wave function at the origin \cite{appelquist75}. In the $m_f \to 0$ limit, we find
\begin{eqnarray}
  \Gamma(^3S_1\rightarrow f\bar{f}) = N_{c}Q_{f}^2\frac{2\pi\alpha^2}{3m_L^2} |\Psi(0)|^2
    \left[ 
    \begin{array}{l}
    \left(1+2\textbf{Re}(\chi_0)g^L_Vg^{f}_V\right)\\
    +\left|\chi_0\right|^2 \left({|g^L_V|}^2+{|g^L_A|}^2\right)\left({|g^{f}_V|}^2+{|g^{f}_A|}^2\right)
    \end{array}
    \right] , \label{3Sdecay1}
\end{eqnarray}
where $\chi_0:=\chi|_{s=4m_L^2}$ with $\chi$ given by Eq.~(\ref{chidef}) [with $Q_q^2$ replaced by $Q_f^2$]. This expression is appropriate for all 
fermions except for the especially heavy top quark. For the top quark we find
\begin{eqnarray}
  \Gamma(^3S_1\rightarrow t\bar{t}) &=& \frac{8\pi\alpha^2}{9m_L^2}|\Psi(0)|^2\sqrt{1-\frac{m_t^2}{m_L^2}} \nonumber \\ 
    &&\times\left[ 
    \begin{array}{l}
    \left(1+2\textbf{Re}(\chi_0)g^L_Vg^{t}_V\right)\left(1+\frac{m_t^2}{2m_L^2}\right)\\
    +\left|\chi_0\right|^2 \left({|g^L_V|}^2+{|g^L_A|}^2\right)\left({|g^{t}_V|}^2+{|g^{t}_A|}^2\right)\left(1-\frac{m_t^2}{4m_L^2}\right)
    \end{array}
    \right] .  \label{3Sdecay2}
\end{eqnarray}

In our analysis the $W^+W^-$, $Z$ Higgs, $hh\gamma$ and $hhZ$ decay channels of the $^3S_1$ state are ignored. 
For lipton masses greater than 100 GeV they each contribute of order $2\%$ \cite{carlson91}. We take the total decay rate as
\begin{eqnarray}
  \Gamma(^3S_1)&=&2\Gamma(^3S_1\rightarrow u\bar{u})+\Gamma(^3S_1\rightarrow t\bar{t})+3\Gamma(^3S_1\rightarrow d\bar{d}) \nonumber \\
    & & + 3\Gamma(^3S_1\rightarrow e^+e^-)+3\Gamma(^3S_1\rightarrow \nu_e\overline{\nu_e}) \label{3s1width}
\end{eqnarray}
and use Eqs.~(\ref{3Sdecay1}) and (\ref{3Sdecay2}) to find the branching fractions. The results are shown in 
Fig.~\ref{branchings}. We include the experimentally excluded region, $m_L<100$ GeV, so that the interested reader may compare the results with those for the colour SU(5) model~\cite{carlson91}.

\subsubsection{Cross section}

\begin{figure}[t]
\includegraphics{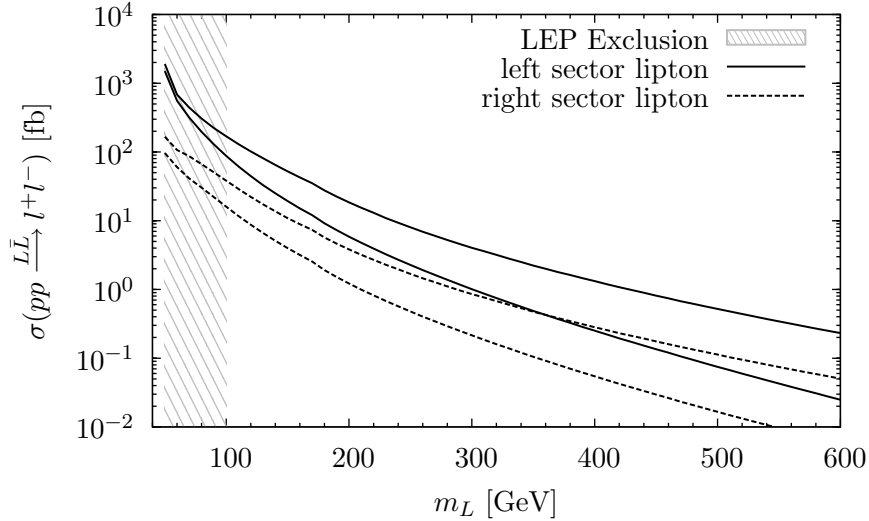}
\caption{\small Cross section of $pp\rightarrow l^+l^-$ through the production of liptonium and the subsequent decay of the $^3S_1$ state for the LHC at 14 TeV and 7 TeV descending.}
\label{hhleptons}
\end{figure}

The cross section of two hadrons to a single lepton pair via the liptonium channel can now be found using Eq.~(\ref{ppLLbarll}). 
The resulting cross sections for the LHC are shown in Fig.~\ref{hhleptons}.

\subsubsection{$^3S_1$ width}

The vertical axis of Fig.~\ref{hhleptons} can be read as `the number of events per 1 fb$^{-1}$ of integrated luminosity in the dilepton channel', 
where `dilepton' could either mean dielectron or dimuon. 
Given enough data, these events would be detected as a peak in the dilepton invariant mass spectrum at $m_{ll} \sim 2m_L$. 
We would like to have some idea of the width of this peak. To achieve this we must evaluate Eq.~(\ref{3s1width}).

With the liptonium in the one huon exchange potential (\ref{onehuonpot}) regime, 
standard results for hydrogenic wave functions tell us that, with fixed $\alpha_s(r)=\alpha_s(m_L)$, we have
\begin{eqnarray}
  |\Psi_n(0)|^2=\left(\frac{3}{8}n^{-1}m_L\alpha_S(m_L)\right)^3/\pi ,
\end{eqnarray}
where $n$ is the radial quantum number. For $n=1$ and $m_L$ varying from 100$-$600 GeV, $\Gamma({^3S_1})$ varies from $\sim1$$-$$100$ keV.

The energy resolution of the ATLAS electromagnetic liquid argon calorimeter is $\sigma(E)/E=10\%/\sqrt{E (\text{GeV})}+0.7\%$ \cite{aurousseau11}. 
The detector sees a convolution of the true peak and the Gaussian defined by variance $\sigma^2(E)$. 
Since the resolution is much larger than the width of the liptonium invariant mass peak, we can assume that the 
observed peak is a Gaussian with $68\%$ of events in the window of width $2\sigma(E)$ surrounding the peak energy $2m_L$.

\subsubsection{Seeing liptons}


\begin{figure}[t]
\includegraphics{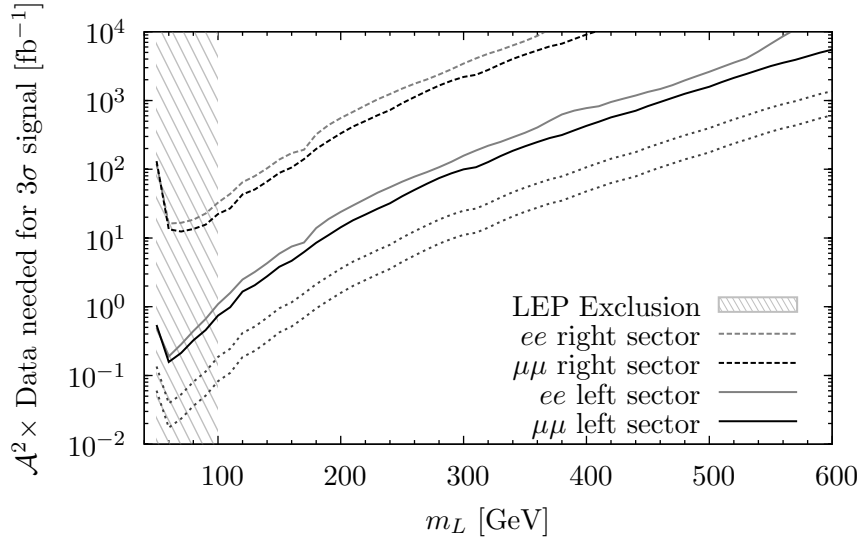}
\caption{\small The amount of data in fb$^{-1}$ we need to see a liptonium resonance at 3$\sigma$ 
significance multiplied by the acceptance squared. [For $m\sigma$ significance multiply by a factor $m^2/9$]. 
The dotted grey lines are for dimuon channel left sector liptons with SU(5) or SU(7) leptonic colour descending.
}
\label{liptonfb}
\end{figure}

We now make an estimate for the amount of data needed to see a liptonium resonance in the dilepton channel. 
For an invariant mass of $m_{ll}$ we employ a bin width of $2\sigma(m_{ll})$. We then compare the number of 
background events, $N_b$, with the number of signal events, $N_s$, in this bin. 

The number of background events in the dimuon and dielectron channels is determined by extrapolating the number of SM events expected in an ATLAS dilepton resonance search \cite{atlaszpnew} using
\begin{eqnarray}
  N=\sigma\int\mathcal{L}dt, \label{NsigL} 
\end{eqnarray}
where $\int\mathcal{L}dt$ is the integrated luminosity. 

The number of signal events is given by $N_s=\mathcal{A}\times\sigma(pp\stackrel{L\bar{L}}{\longrightarrow}l^+l^-)\int\mathcal{L}dt$, 
where $\mathcal{A}$ is the acceptance for the signal assuming the same analysis cuts as in Ref.~\cite{atlaszpnew}.
The typical value for a spin-1 resonance with these cuts is $\mathcal{A}\approx0.65$ ($\mathcal{A}\approx0.40$) in the dielectron (dimuon) channel.

The number of observed background events fluctuates according to a Poisson distribution with a standard deviation of $\sqrt{N_b}$. 
If we know the number of background and the number of signal events for 1 fb$^{-1}$ of data, 
$n_b$ and $n_s$, we can extrapolate to the number of events in $x$ fb$^{-1}$ of data using Eq. (\ref{NsigL}): $xn_b$ and $xn_s$. 
The point where the signal reaches 3$\sigma$ significance inside the window, or when $N_s=3\sqrt{N_b}$, is at
\begin{eqnarray}
  x = \frac{9n_b}{{n_s}^2} . \label{invfbneeded}
\end{eqnarray}
This is the point at which the liptonium resonance is noticeable.

The results versus lipton mass are shown in Fig.~\ref{liptonfb}. Also shown in Fig.~\ref{liptonfb} are the cases of the closely related 
SU(5) and SU(7) leptonic colour models \cite{foot06}. The cross section to diliptons is proportional to a colour factor 
$N-1$ for SU($N$) leptonic colour, so, compared to the SU(3) leptonic colour model, the amount of integrated 
luminosity needed for 3$\sigma$ signal significance decreases by $1/4$ and $1/9$ for SU(5) and SU(7) respectively.

These calculations have been carried out for the LHC running at 7 TeV. They can readily be 
recalculated for the LHC at 14 TeV following a similar procedure. We stress that this analysis is simplified and could be improved with smarter analysis cuts. 
Our results illustrate that in the foreseeable future, with the LHC set to gather more than $100$ fb$^{-1}$ of data, 
the LHC will explore lipton mass parameter space beyond $100$ GeV, an improvement on the LEP frontier.

A similar calculation for the diphoton product of the liptonium $^1S$ state shows that we need $\sim10^6$ fb$^{-1}$ to see a 100 GeV liptonium resonance at 3$\sigma$ significance. 
A liptonium dilepton resonance will therefore present itself with the absence of such a resonance in the diphoton channel. 
This is one way of discerning the resonance from other particles such as the Randall-Sundrum graviton \cite{rscms,*rsatlas}.

\section{\label{sec5}Colour triplet diquark}

We will now focus on the phenomenology of the Higgs scalars contained within the Yukawa Lagrangian (\ref{Yuk2}). We are interested in signatures at the LHC, 
so we concentrate on the QCD coloured scalar $\chi_2$, a colour triplet diquark. Diquarks are 
generically predicted by a number of bottom-up models and GUTs, and are well studied in the literature \cite[see e.g.\ Ref.][]{giudice11}.

\subsection{Constraints}

Rewriting the $\chi_2$ Lagrangian with generational indices we have
\begin{eqnarray}
  \mathcal{L}= 2(h_1)_{ij}\overline{(u_L^i)^c}d_L^j\chi_2 + 
                (h_2)_{ij}\overline{(u_R^i)^c}d_R^j\chi_2 + H.c.
\end{eqnarray}
Colour asymmetry and asymmetry in the SU(2) contraction implies that $h_1$ is symmetric in flavour space when the quarks are in the gauge basis. 
Flavour rotation to the mass basis does not retain this symmetry, because distinct rotations must be 
applied to $u_L$ and $d_L$. There is no symmetry/antisymmetry condition on the $h_2$ matrix. 

The $\chi_2$ diquark transforms as $(3,1)$ under SU(3)$_q\times$SU(2)$_L$. If one considers only these SM charges then the scalar can also behave as a leptoquark through the terms 
$\overline{(f_L)^c}q_L\overline{\chi}$ and $\overline{(e_R)^c}u_R\overline{\chi}$. These terms lead to rapid decay of the proton 
through tree level exchange unless the mass of the scalar is very large, so they need to be forbidden if we are to have any 
chance of seeing this scalar at the LHC. The usual mechanism is baryon parity \cite{giudice11} or U(1) symmetry \cite{ajaib09}. 
The q-l symmetric model provides a natural mechanism. If one tries to write down the terms 
$\overline{(F_L)^c}Q_L\overline{\chi}$ and $\overline{(u_R)^c}E_R\overline{\chi}$ it quickly becomes clear that they are not gauge 
invariant unless $\chi$ transforms as a 3 under both SU(3)$_q$ and SU(3)$_l$. So this $\chi$ cannot be our $\chi_2$. 
Further, if we tried to replace $\chi_2$ with such a $\chi$ [transforming like $(3,3,1)$] the terms 
$\overline{(Q_L)^c}Q_L\chi_2$ and $\overline{(u_R)^c}d_R\chi_2$ would not be gauge invariant under SU(3)$_l$. 
So q-l symmetry naturally forbids the leptoquark terms. Thus there are no constraints on $\chi_2$ from proton decay.


We must also suppress flavour changing neutral currents (FCNCs) to avoid strong constraints from meson-antimeson mass mixing measurements. 
To achieve this we must adopt for $h_1$ and $h_2$ a suitable Yukawa-coupling ansatz. It will be sufficient for us to 
define the $h_1$ and $h_2$ matrices diagonally in the mass basis, as in Ref.~\cite{gogoladze10}. In this way we avoid FCNC concerns.

\subsection{Cross section}

\begin{figure}[t]
	\centering
		\scalebox{0.75}{\includegraphics{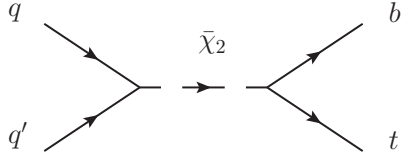}}
	\caption{\small Single top production mediated by the $\chi_2$ diquark.}
	\label{chi2}
\end{figure}

We now turn to phenomenology at hadron colliders. If we restrict ourselves to considering proton valence quarks, 
with a diagonal Yukawa-coupling ansatz we are considering only three parton-level processes: the $s$-channel processes $ud\rightarrow ud, cs, tb$. 
The first two processes result in a dijet final state, whilst the last is a single top + $b$-jet. The single top signal is the preferred method for a $\chi_2$ diquark 
search at the LHC due to a lower SM background \cite{cmsscalardijet,*atlasscalardijet}. The $s$-channel process mediated by $\chi_2$ and producing a single top is shown in Fig.~\ref{chi2}.

Our goal is to calculate $\sigma(pp\stackrel{\bar{\chi}_2}{\longrightarrow}tb)$ as a function of $\chi_2$ mass and couplings. To facilitate our calculation we write the Lagrangian in the Dirac basis:
\begin{eqnarray}
  \mathcal{L}= \overline{(u^i)^c}[(g_V)_{ij}-(g_A)_{ij}\gamma_5]d^j\chi_2+H.c. ,
\end{eqnarray}
where
\begin{eqnarray}
\begin{array}{lcl}
  (g_V)_{ij}&=&\frac{1}{2}[2(h_1)_{ij}+(h_2)_{ij}], \\
  (g_A)_{ij}&=&\frac{1}{2}[2(h_1)_{ij}-(h_2)_{ij}] .
\end{array}
\end{eqnarray}
Because $h_1$ and $h_2$ are diagonal, $g_V$ and $g_A$ are also diagonal.

The dominant parton level process is $ud\stackrel{\bar{\chi}_2}{\longrightarrow}tb$. The cross section in the $m_{u,d,b}\ll m_t$ limit is
\begin{eqnarray}
  \sigma(ud\stackrel{\bar{\chi}_2}{\longrightarrow}tb) &=& 
   \frac{1}{12\pi}\left(\frac{s-m_t^2}{s}\right)^2\frac{s}{(s-M_\chi^2)^2+M_\chi^2\Gamma_\chi^2} \nonumber \\  
    &&\times\left(\left|(g_V)_{11}\right|^2+\left|(g_A)_{11}\right|^2\right)\left(\left|(g_V)_{33}\right|^2+\left|(g_A)_{33}\right|^2\right),
\end{eqnarray}
where $M_\chi, \Gamma_\chi$ are the mass and width of $\chi_2$.
The full width is given by the sum of partial widths
\begin{eqnarray}
  \Gamma(\bar{\chi}_2\rightarrow u^{i=[1,2]}d^j) &=& \frac{M_\chi}{4\pi} 
    \left(\left|(g_V)_{ij}\right|^2+\left|(g_A)_{ij}\right|^2\right) , \nonumber \\
  \Gamma(\bar{\chi}_2\rightarrow td^j) &=& \frac{M_\chi}{4\pi}\left(1-\frac{m_t^2}{M_\chi^2}\right)^2 \left(\left|(g_V)_{3j}\right|^2+\left|(g_A)_{3j}\right|^2\right) .
\label{chiwidth}
\end{eqnarray}

The cross section at a hadron collider is expressed in terms of PDFs in a similar way to the Drell-Yan process (\ref{drellyan}): 
\begin{eqnarray}
	\sigma(pp\stackrel{\bar{\chi}_2}{\longrightarrow}tb) &=& \sum_{i,j}\int_0^1\int_0^1dx_1dx_2 
	\left[
	\begin{array}{l}
					f_{u^i/p}(x_1,Q^2)f_{d^j/p}(x_2,Q^2) \\
					+ f_{u^i/p}(x_2,Q^2)f_{d^j/p}(x_1,Q^2)
  \end{array}
  \right] \nonumber \\ 
  &&\times \sigma(u^id^j\stackrel{\bar{\chi}_2}{\longrightarrow}tb;s=x_1x_2S_{beam}). \label{hhchi}
\end{eqnarray}
The cross section $\sigma(pp\stackrel{\chi_2}{\longrightarrow}\bar{t}\bar{b})$ is found similarly. 
At the LHC, top quark production is preferred over anti-top production due to non-zero initial baryon number, so that 
$\sigma(pp\stackrel{\bar{\chi}_2}{\longrightarrow}tb)\gg\sigma(pp\stackrel{\chi_2}{\longrightarrow}\bar{t}\bar{b})$. The case for the Tevatron is different, 
with $\sigma(p\bar{p}\stackrel{\bar{\chi}_2}{\longrightarrow}tb)=\sigma(p\bar{p}\stackrel{\chi_2}{\longrightarrow}\bar{t}\bar{b})$.

The parton level cross section in Eq.~(\ref{hhchi}) will have a distinct peak at $x_1x_2S_{beam}=M_\chi^2$. The height and width of this peak is 
determined by the $\chi_2$ width (\ref{chiwidth}), which can be quite small depending on the coupling. 
This causes numerical problems when integrating Eq.~(\ref{hhchi}) directly. It is convenient instead to introduce a change of variable, $x_2=M_{inv}^2/x_1S_{beam}$, 
so that the peak is only dependent on one integration variable: the invariant mass of $tb$, $M_{inv}$. 
Making this substitution, and only considering valence quarks at the parton level, we obtain
\begin{eqnarray}
	\frac{d \sigma(pp \stackrel{\bar{\chi}_2}{\longrightarrow}tb)}{dM_{inv}} &=& \int_{\frac{M_{inv}^2}{S_{beam}}}^1dx\frac{2M_{inv}}{xS_{beam}}
	\left[
	\begin{array}{l}
					f_{u/p}(x,Q^2)f_{d/p}\left(\frac{M_{inv}^2}{xS_{beam}},Q^2\right) \\
					+ f_{u/p}\left(\frac{M_{inv}^2}{xS_{beam}},Q^2\right)f_{d/p}(x,Q^2)
  \end{array}
  \right] \nonumber \\ 
  &&\times\sigma(ud\stackrel{\bar{\chi}_2}{\longrightarrow}tb;s=M_{inv}^2) . \label{tbcrosssec}
\end{eqnarray}
One may then find the total cross section by integrating over the invariant mass. The total integration is over 
$M_{inv}=(0,\sqrt{S_{beam}})$, with the region near $M_{inv}=M_\chi$ dominating the cross section.

\subsection{Mass/coupling limits}

%

A Bayesian approach to calculating and combining CDF and D0 Tevatron results for the single top cross section is 
presented in Ref.~\cite{heinson11}. 
The quantity includes both $s$-channel and $t$-channel contributions to single top and single anti-top quark final states. 
(In the language of hadron colliders, `single top quark' means single top quark or single anti-top quark.) 
The single top cross section is found to be
\begin{equation}
  \sigma^{\text{Tevatron}}_{(s+t)-ch}(p\bar{p}\rightarrow tX;\bar{t}X)=2.76^{+0.56}_{-0.47} \text{ pb} \label{singletlim}
\end{equation}
for a 170 GeV top quark. This is in good agreement with the SM theoretical prediction of $3.33^{+0.08}_{-0.10}$ pb \cite{kidonakis10,*kidonakis11b}. We can use 
this measurement to put a limit on the cross section of our diquark mediated process. A conservative bound is
\begin{equation}
  \sigma^{\text{Tevatron}}(p\bar{p}\longrightarrow tb;\bar{t}\bar{b}) < 1 \text{ pb} . \label{singlettevlimit}
\end{equation}

For the LHC at 7 TeV, the SM theoretical prediction for the single top $s$-channel cross section is $\sim4.6$ pb \cite{kidonakis10}. 
There is not yet enough data to statistically observe single top production in this channel. However, the preliminary limit \cite{deroeck},
\begin{equation}
  \sigma^{\text{LHC@7TeV}}(pp\longrightarrow tX;\bar{t}X) < 26.5 \text{ pb} , \label{singletatlaslimit}
\end{equation}
is enough to compete with Tevatron bounds, as we will see.

\begin{figure}[t]
  \includegraphics{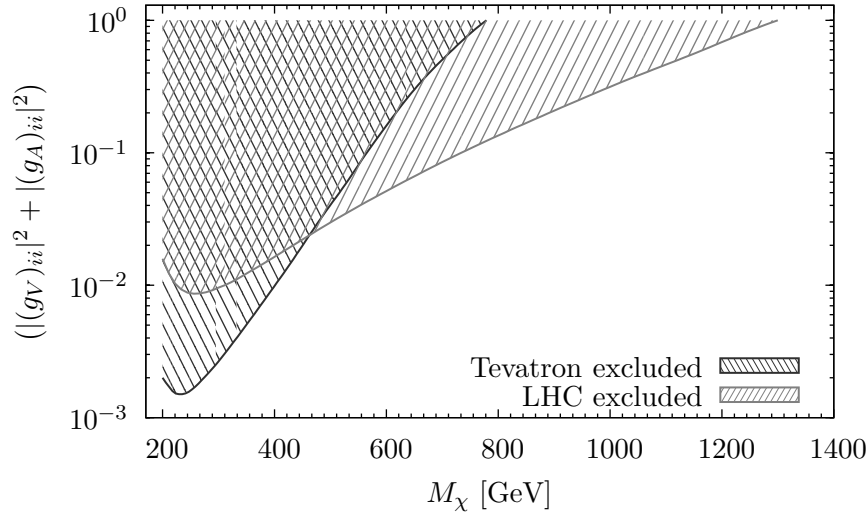}
  \caption{\label{figchispace} The experimentally excluded regions of $\chi_2$ parameter space.}
\end{figure}

We calculated $\sigma(pp\rightarrow tb; \bar{t}\bar{b})$ and $\sigma(p\bar{p}\rightarrow tb; \bar{t}\bar{b})$ as a function of 
$M_\chi$ and $\left(|(g_V)_{ii}|^2+|(g_A)_{ii}|^2\right)$ using Eq.~(\ref{tbcrosssec}). Here we made no assumptions on the 
magnitude relationship between entries of the generational matrices $h_1$ and $h_2$. PDFs were evaluated using the MSTW08 PDF set at $Q^2=\frac{1}{4}s$. 
A $k$-factor of 1.3 was used for both the Tevatron and LHC calculations \cite{han10}. 
We then used Eqs.~(\ref{singlettevlimit}) and (\ref{singletatlaslimit}) to put constraints on the two dimensional mass-coupling space. 
Our results are shown in Fig.~\ref{figchispace}. We see that the LHC can already impose bounds that exceed those of the Tevatron at sufficiently large coupling.

\section{\label{sec6}Conclusion}

We have studied the phenomenology of the quark-lepton symmetric model of Foot and Lew \cite{foot90} in the context of the LHC. 
In this model an SU(3)$_l$ `leptonic colour' symmetry is appended to the SM gauge group, tripling the number of SM leptons. 
Consequently a $Z_2$ discrete quark-lepton symmetry of the Lagrangian can be defined. 
The additional symmetries are then spontaneously broken by an SU(3)$_l$ triplet Higgs multiplet, 
the $Z_2$ partner of which is an SU(3)$_q$ triplet diquark called $\chi_2$. 
After this breaking, the eight gauge bosons of SU(3)$_l$ become a heavy neutral $Z'$ gauge boson [in conjunction with the
gauge boson of U(1)$_X$], three massless exotic gluons (huons) mediating an unbroken confining SU(2)$'$ force, and four charge $\pm1/2$ `weak' bosons [these four bosons do not couple to quarks and are not expected to exhibit prominent LHC phenomenology]. There are also four massive $\pm1/2$ charged exotic leptons (liptons) per family which transform as doublets under SU(2)$'$. The SM particles are uncharged under SU(2)$'$.

We investigated three main areas of interest at the LHC: the Z$'$ heavy neutral gauge boson, the liptons, and the diquark $\chi_2$. 

Using dilepton data, we put lower bounds on the mass of the Z$'$ gauge boson of 
$780$ GeV and $950$ GeV, from the the Tevatron and LHC, respectively. The latter is an improvement of 230 GeV 
on the previous best limit \cite{foot91}.

We then considered lipton pair production, $L\bar{L}$, at the LHC through the Drell-Yan mechanism. 
All pairs form excited `liptonium' bound states. These states then deexcite via isotropic hueball and soft photon radiation, 
resulting in large missing energy. It was estimated that $75\%$ of liptonium states will annihilate to difermions whilst 25$\%$ 
annihilate to diphotons and huons. The diphoton events were found to be too few in number to detect at the LHC. 
However, we found that annihilation to dileptons will be observable at the LHC in the near future for lipton mass $m_L>100$ GeV, 
the current bound set by the LEP experiments. New lipton mass parameter space is already being explored by the LHC 
for the closely related SU(5) and SU(7) leptonic colour models.

Lastly, we investigated the phenomenology of the colour triplet scalar diquark $\chi_2$. Using single top production data from the
Tevatron and LHC we excluded areas of the two dimensional mass-coupling parameter space, as shown in Fig.~\ref{figchispace}. 
We found that the LHC is already competitive with Tevatron bounds.

\acknowledgments

This work was supported in part by the Australian Research Council.

\bibliographystyle{apsrev4-1}
\bibliography{references2}

\end{document}